\documentclass[aps,preprintnumbers,nofootinbib,superscriptaddress,10pt]{revtex4}
\usepackage[dvipdfmx]{graphicx} 
\usepackage{amsmath,amssymb,amsfonts,bm,cancel} 

\def\a{\alpha}    \def\d{\delta}     \def\th{\theta}   \def\l{\lambda}                  

\def\dg{\dagger}  \def\nn{\nonumber}

\newcommand{\lsp}{  ( } \newcommand{\rsp}{  ) } \newcommand{\Lg}{\mathcal{L}}

\def\abs#1{| #1|}

\newcommand{\row}[2]{ \begin{pmatrix}  #1 & #2   \end{pmatrix}  }
\newcommand{\column}[2]{ \begin{pmatrix}  #1 \\ #2 \\  \end{pmatrix} }

\newcommand{\Diag}[3]{ \begin{pmatrix} #1 & 0 & 0 \\ 0 & #2 & 0 \\ 0 & 0 & #3 \\\end{pmatrix}}

\begin{document}

%%%%%%%%%%%%%%%%%%%%%%%%%%%%%%%%%%%%%%%%

\title{\large Perturbative determination of quark CP phase and \\ rephasing invariants of hierarchical mass matrices using their inverses }

\preprint{STUPP-26-300}
%%%%%%%%%%%%%%%%%%%%%%%%%%%%%%%%%%%%%%%%
\author{Masaki J. S. Yang}
\email{mjsyang@mail.saitama-u.ac.jp}
\affiliation{Department of Physics, Saitama University, 
Shimo-okubo, Sakura-ku, Saitama, 338-8570, Japan}
%\affiliation{Department of Physics, Graduate School of Engineering Science, Yokohama National University, Yokohama, 240-8501, Japan}

%%%%%%%%%%%%%%%%%%%%%%%%%%%%%%%%%%%%%%%%

%\date{\today}

%%%%%%%%%%%%%%%%%%%%%%%%%%%%%
\begin{abstract} %%%%%%%%%%%%%%%%%%%%%
%%%%%%%%%%%%%%%%%%%%%%%%%%%%%

In this paper, we derive approximate expressions for the CP phase $\delta$ in the CKM matrix 
by a perturbative singular value decomposition for hierarchical mass matrices of quarks $m_{u,d}$.
The diagonalization is achieved through a seesaw-like procedure in which the heavier generations are successively integrated out, naturally leading to mixing matrices expressed in terms of the mass matrices and their inverses  $m_{u,d}^{-1}$. 
As a result, in the basis where the up-type quark mass matrix is diagonal, $\delta$ is reduced to a  fourth-order invariant $\delta \simeq \arg [ - m^{-1}_{d12} m_{d23}^{} / m^{-1}_{d11} m_{d13}^{} ]$.
Furthermore, in the Kobayashi--Maskawa parametrization, the CP phase is likewise expressed by an invariant
$\delta_{\rm KM} \simeq \arg [ m^{-1}_{d13} m_{d33}^{} / m^{-1}_{d11} m_{d13}^{} ] \simeq \pi/2$ in the same basis. 
Since the next-to-leading-order corrections are suppressed at second order in the perturbation, 
the accuracy of these expressions is typically of order $O(\l^{2}) \sim 4 \%$.

%%%%%%%%%%%%%%%%%%%%%%%%%%%%%
\end{abstract} %%%%%%%%%%%%%%%%%%%%%%
%%%%%%%%%%%%%%%%%%%%%%%%%%%%%

\maketitle

%%%%%%%%%%%%%%%%%%%%%%%%%
\section{Introduction}
%%%%%%%%%%%%%%%%%%%%%%%%%

CP violation has been one of the central issues in particle physics. Understanding its mathematical structure is a key ingredient toward uncovering the origin of this symmetry breaking. 
In the Standard Model, the CP violation in the quark sector is described by a single phase contained in the Cabibbo--Kobayashi--Maskawa (CKM) matrix \cite{Cabibbo:1963yz, Kobayashi:1973fv}. 
Although the mixing matrix depends on a choice of parameterizations, 
physical observables are formulated in terms of rephasing invariants 
\cite{Wu:1985ea,Bernabeu:1986fc,Gronau:1986xb,Branco:1987mj,Bjorken:1987tr,Nieves:1987pp,Botella:1994cs,Kuo:2005pf, Farzan:2006vj, Jenkins:2007ip,Branco:2008ai, Chiu:2015ega}. 

Traditionally, the Jarlskog invariant $J$ \cite{Jarlskog:1985ht} has played a central role in the description of CP violation. 
Since the invariant $J$ is a small quantity of order $O(10^{-5})$, 
it is highly sensitive to various approximations. 
Capturing analytic behavior of the CP violation---including error estimation---has been a technically challenging task due to the computational precision.
Although Ref.~\cite{Hall:1993ni} presented a general perturbative treatment of the CKM matrix and its CP phase, such a general treatment has not been pursued in later literature.

In recent years, rephasing invariants involving the determinant of the mixing matrix
have been proposed for a direct evaluation of the CP phase 
\cite{Yang:2025hex,Yang:2025law,Yang:2025ftl,Yang:2025vrs, Yang:2025cya,Yang:2025dhm}. 
It complements conventional approaches in both analytic structure and error estimation.
In subsequent studies \cite{Yang:2025dkm, Yang:2025qlg, Yang:2026ulu, Yang:2026wjg, Yang:2026pdm, Yang:2026qjf}, various rephasing transformations to different parameterizations, which had been discussed only at a conceptual level, were displayed explicitly  by employing the determinant as a carrier of global phase information. 
However, these studies have primarily focused on the mixing matrix, 
leaving relationships between magnitudes of CP phases and mass matrices not fully understood. 
Although it is well known from Jarlskog's work that the magnitude of CP violation is related to the commutator of the quark mass matrices, the imaginary part of this invariant does not isolate the behavior of the CP phase itself. 

In this paper, we derive a general perturbative expression for the CP phase in the CKM matrix 
for hierarchical mass matrices. 
By successively integrating out heavier generations through a seesaw-like procedure, 
the diagonalization of mass matrices $m_{u,d}$ is naturally described in terms of the mass matrices and their inverses  $m_{u,d}^{-1}$. 
Using rephasing invariant formulae, the CP phases  in the Particle Data Group (PDG) and Kobayashi--Maskawa (KM) parametrizations can be expressed in terms of the mixing matrix. 
In the basis where the up-type quark mass $m_{u}$ is diagonal, these phases are given by the fourth-order rephasing invariant which is constructed from the down-quark mass matrix $m_{d}$ and its inverse $m_{d}^{-1}$.
We also systematically evaluate the next-to-leading-order (NLO) corrections and show that they are suppressed by the second order of the Cabibbo angle $\l^{2} \sim O(4 \%)$ relative to the leading-order (LO) contributions in most cases.
Consequently, for hierarchical quark mass matrices, the LO expressions already provide an accuracy comparable to the current experimental precision. 

This paper is organized as follows. 
The next section gives a formulation of the perturbative singular value decomposition. 
In Sec.~3, we discuss the CP phase represented by rephasing invariants using the mass matrices 
and their inverses. 
The final section is devoted to conclusions. 

%%%%%%%%%%%%%%%%%%%%%%%%%
\section{Perturbative singular value decomposition of fermion mass matrices}
%%%%%%%%%%%%%%%%%%%%%%%%%

We formulate the perturbative singular value decomposition (SVD) applicable to generic hierarchical mass matrices.
Such an approximate diagonalization has found widespread application in the study of the right-handed neutrino mass matrix \cite{Akhmedov:2003dg}, particularly in analyses of SO(10)-inspired $N_2$ leptogenesis 
\cite{DiBari:2005st, Vives:2005ra, DiBari:2008mp, Bertuzzo:2010et,DiBari:2010ux, DiBari:2013qja, DiBari:2014eqa, DiBari:2014eya, DiBari:2017uka, Chianese:2018rnq, DiBari:2021fhs}. 
Here, we extend this method to general mass matrices that are not necessarily symmetric.

We first define a mass matrix $m$ to be {\it hierarchical} 
if the magnitudes of certain off-diagonal elements of $m$ are sufficiently smaller than those of the adjacent matrix elements of the heavier generations 
\begin{align}
|m_{33}| \gg |m_{23}| , |m_{32}| \gg |m_{13}| , |m_{31}| \, , 
~~~ 
|m_{22}| \gg |m_{21}| , |m_{12}| \, . 
\label{1}
\end{align}
When these conditions are satisfied, the mass matrix $m$ approximately possesses chiral flavor symmetries $U(1)_L^2 \times U(1)_R^2$ 
\begin{align}
m' 
= \Diag{e^{i \a_{1 L}}}{e^{i \a_{2 L}}}{1} 
m
\Diag{e^{i \a_{1 R}}}{e^{i \a_{2 R}}}{1} 
\simeq m \, . 
\end{align}
Defining the SVD of the matrix $m$ by $U_L\, m\, U_R^\dagger = m^{\rm diag}$ with  the diagonal mass matrix $m^{\rm diag}$, the unitary matrices $U_L$ and $U_R$ necessarily involve only small mixings. 

Successively integrating out the heavier generations,
we determine the singular values using a seesaw-like procedure. 
The validity of this approach will be justified more rigorously later.
Integrating out the heaviest third generation $f_{(L,R)3}$, the resulting mass matrix for the remaining two lighter  generations $m_0$ is
\begin{align}
m_{0} 
 & = 
\begin{pmatrix}
m_{11} & m_{12} \\
m_{12} & m_{22} 
\end{pmatrix} 
 - {1\over m_{33} } 
\column{ m_{13} }{ m_{23} } \otimes
\row{ m_{13} }{ m_{23} } 
 = {\det m \over m_{33}}
\begin{pmatrix}
 m^{-1}_{22} & -m^{-1}_{12} \\
 -m^{-1}_{21} & m^{-1}_{11} \\
\end{pmatrix} . 
\label{3}
\end{align}
Here, $m^{-1}_{ij}$ denotes the $(i,j)$ element of the inverse $m^{-1}$. 
Integrating out the second generation $f_{(L,R)2}$ once more, the mass matrix is reduced to one  generation, yielding the diagonal matrix $m^{\rm diag}$ as
\begin{align}
m^{\rm diag} = {\rm diag} \, ({1 \over m_{11}^{-1} } \, ,  {  m_{11}^{-1} \det m \over m_{33}  } \, ,  m_{33} ) \, . 
 \label{diag}
\end{align}
For realistic charged fermions, the following hierarchy is required 
\begin{align}
{1 \over | m_{11}^{-1}| } \ll  \left | { m_{11}^{-1} \det m \over m_{33}  } \right |  \ll | m_{33} | \, . 
\label{hierarchy}
\end{align}

As the heavy states are successively integrated out, the perturbative diagonalization is achieved as
\begin{align}
U_{L2} U_{L 1 } m U_{R 1}^{\dg} U_{R2}^{\dg} \simeq m^{\rm diag} \, , 
\end{align}
where
\begin{align}
 U_{L 1} &= 
\begin{pmatrix}
 1 & 0 & -\frac{m_{13}}{m_{33}} \\
 0 & 1 & -\frac{m_{23}}{m_{33}} \\
 \frac{m_{13}^*}{m_{33}^*} & \frac{m_{23}^*}{m_{33}^*} & 1 \\
\end{pmatrix} , 
~~~
U_{L2} = 
\begin{pmatrix}
 1 & \frac{m_{12}^{-1}}{m_{11}^{-1}} & 0 \\
- \big( \frac{m_{12}^{-1}}{m_{11}^{-1} } \big)^{*} & 1 & 0 \\
 0 & 0 & 1 \\
\end{pmatrix} ,  \\
U_{R 1}^{\dg} &=
\begin{pmatrix}
 1 & 0 & \frac{m_{31}^*}{m_{33}^*} \\
 0 & 1 & \frac{m_{32}^*}{m_{33}^*} \\
 -\frac{m_{31}}{m_{33}} & -\frac{m_{32}}{m_{33}} & 1 \\
\end{pmatrix}  , 
~~~
U_{R2}^{\dg} 
= 
\begin{pmatrix}
 1 & - \big( \frac{ m_{21}^{-1} }{ m_{11}^{-1}} \big)^{*} & 0 \\
 \frac{m_{21}^{-1}}{m_{11}^{-1} } & 1 & 0 \\
 0 & 0 & 1 \\
\end{pmatrix} . 
\label{UR}
\end{align}
The perturbative treatment of the diagonalization is valid when the following conditions are satisfied 
\begin{align}
|{m_{i3}}/{m_{33}}| , |{m_{3j}}/{m_{33}}| \lesssim 0.1 \, , ~~~ 
|m_{12}^{-1}/ m_{11}^{-1}| \, ,  |m_{21}^{-1} / m_{11}^{-1} | \lesssim 0.1 . 
\end{align}
In the mixing between the first and second generations of quarks, however, this upper bound may be as large as approximately $0.2$ in order to reproduce the Cabibbo angle.
For sufficiently small mixing in the diagonalization of a hierarchical matrix, 
 its inverse matrix $m^{-1} \simeq  U_{R 1}^{\dg} U_{R2}^{\dg} (m^{\rm diag})^{-1} U_{L2} U_{L1}$ is likewise inversely hierarchical and satisfy 
\begin{align}
|m_{33}^{-1}| \ll |m_{23}^{-1}| , |m_{32}^{-1}| \ll |m_{22}^{-1}| \, , 
~~~ 
|m_{21}^{-1}| , |m_{12}^{-1}| \ll |m_{11}^{-1}| \, . 
\end{align}
Therefore, these perturbative conditions are justified in many cases.
In what follows, these inequalities are assumed to hold.

%%%%%%%%%%%%%%%%%%%%%%%
\subsection{Explicit Perturbative Expansion}
%%%%%%%%%%%%%%%%%%%%%%%

We demonstrate that the seesaw-like procedure indeed constitutes the perturbative SVD.
The Hermitian matrix $m m^\dagger$ is manifestly invariant under the right-handed field redefinition
$m' = m V_R$. We decompose it as follows:
\begin{align}
m m^{\dg} = H^{1} + H^{2} + H^{3} , 
\end{align}
where $H^{i}_{jk} = m_{ji} m_{k i}^{*}$ and 
\begin{align}
H^{1} & = 
\begin{pmatrix}
 m_{11} m_{11}^* & m_{11} m_{21}^* & m_{11} m_{31}^* \\
 m_{21} m_{11}^* & m_{21} m_{21}^* & m_{21} m_{31}^* \\
 m_{31} m_{11}^* & m_{31} m_{21}^* & m_{31} m_{31}^* \\
\end{pmatrix} ,  ~~~
H^{2}  =
\begin{pmatrix}
 m_{12} m_{12}^* & m_{12} m_{22}^* & m_{12} m_{32}^* \\
 m_{22} m_{12}^* & m_{22} m_{22}^* & m_{22} m_{32}^* \\
 m_{32} m_{12}^* & m_{32} m_{22}^* & m_{32} m_{32}^* \\
\end{pmatrix} ,  \nn \\
H^{3} & =
\begin{pmatrix}
 m_{13} m_{13}^* & m_{13} m_{23}^* & m_{13} m_{33}^* \\
 m_{23} m_{13}^* & m_{23} m_{23}^* & m_{23} m_{33}^* \\
 m_{33} m_{13}^* & m_{33} m_{23}^* & m_{33} m_{33}^* \\
\end{pmatrix} . 
\end{align}
In this form, $H^{3}$ provides the leading contribution for a hierarchical mass matrix, and the diagonalization by $U_{L1}$ is exact 
\begin{align}
U_{L1} H^{3} U_{L1}^{\dg} 
=  \Diag{0}{0}{  { |( m^{\dg} m)_{33}  /   m_{33} |^2} } . 
\label{m3L1}
\end{align}

Similarly, diagonalization by $U_{L2} U_{L1}$ eliminates the first row and column of $H^{2}$ 
\begin{align}
U_{L2} U_{L1} H^{2} U_{L1}^{\dg} U_{L2}^{\dg}  = 
\begin{pmatrix}
 0 & 0 & 0 \\[3pt]
 0 & 
 \abs {\frac{\det m }{ m^{-1}_{11} m_{33} } }^{2} (| m^{-1}_{11}| ^2+| m^{-1}_{12}| ^2)^{2}  &
 \frac{\det m (m^{\dg} m)_{23}  }{(m^{-1}_{11})^* m_{33}^2 }  (| m^{-1}_{11}| ^2+| m^{-1}_{12}| ^2) \\[3pt]
 0 &
 * & 
{ | (m^{\dg} m)_{23}/ m_{33} | ^2} \\
\end{pmatrix} . 
\label{m2L1L2}
\end{align}
Note that this is a rank-one Hermitian matrix and can therefore be written as a tensor product of a vector. 
The matrix element denoted by $*$ is fixed by Hermiticity as the complex conjugate of the corresponding transposed element.

Using the perturbativity condition $|m^{-1}_{11}| \gg |m^{-1}_{12}|$, it reduces to
\begin{align}
U_{L2} U_{L1} H^{2} U_{L1}^{\dg} U_{L2}^{\dg} 
\simeq 
\begin{pmatrix}
 0 & 0 & 0 \\[3pt]
 0 &
\abs { \frac{ m_{11}^{-1} \det m }{ m_{33} }}^{2} &
   \frac{(m_{11}^{-1} \det m ) }{m_{33}^2} ( m^{\dg} m )_{23} \\[3pt]
 0 &
 * & 
{ | (m^{\dg} m)_{23}/ m_{33} | {}^2} \\
\end{pmatrix} . 
\end{align}
Since off-diagonal elements do not contribute to the eigenvalues at first-order perturbation, the second singular value is consistent with Eq.~(\ref{diag}).

Finally, in this basis, $H^{1}$ becomes 
\begin{align}
& U_{L2} U_{L1} H^{1} U_{L1}^{\dg} U_{L2}^{\dg} \nn \\
& =
\begin{pmatrix}
 \frac{1}{| m^{-1}_{11}| ^2} & 
 { \det m^{*} \over m^{*}_{33}}  \frac{m^{-1}_{11} (m^{-1}_{21})^*+m^{-1}_{12} (m^{-1}_{22})^*}  {(m^{-1})_{11}^2 } & 
 \frac{  (m^{\dg} m)_{13}}{m_{33} m^{-1}_{11}}  \\[4pt]
* &
 \big | { \det m \over m_{33}} \big |^{2} \frac{ | m^{-1}_{11} (m^{-1}_{21})^*+m^{-1}_{12} (m^{-1}_{22})^* |^{2} }
{  | m^{-1}_{11}| ^2  } &
- {( m^{\dg} m)_{13} \over m_{33} }{\det m^{*} \over m_{33} } \frac{ ( m_{11}^{-1}  )^* m_{21}^{-1} +  ( m^{-1}_{12})^* m^{-1}_{22}  }{ ( m^{-1}_{11} )^*} \\[4pt]
* &
 * & 
 { |(m^{\dg} m )_{13} /  m_{33} |^2} \\
\end{pmatrix} . 
\label{m1L1L2}
\end{align}
Since the off-diagonal elements of $H^{1,2}$ are suppressed by $m_{33}$, they do not contribute at leading order,  
and all the singular values at leading order agree with Eq.~(\ref{diag}).

As physical observables, the left-handed diagonalization matrix $U_L \equiv  U_{L2}U_{L1}$ is the relevant quantity. Combining the two diagonalization matrices, we obtain
\begin{align}
U_{L} = 
U_{L2}  U_{L 1} & \simeq  
\begin{pmatrix}
 1 & \frac{m^{-1}_{12}}{m^{-1}_{11}} & \frac{m^{-1}_{13}}{m^{-1}_{11}} \\[4pt]
- \big( \frac{m^{-1}_{12}}{m^{-1}_{11}} \big)^{*} & 1 & -\frac{m_{23}}{m_{33}} \\[4pt]
 \frac{m_{13}^*}{m_{33}^*} & \frac{m_{23}^*}{m_{33}^*} & 1 \\
\end{pmatrix} .  \label{21}
\end{align}
The (1,3) element of $U_{L}$ is not a misprint, but due to the sum of two terms as 
\begin{align}
 -\frac{m_{13}}{m_{33}} -\frac{m_{23}}{m_{33}}  \frac{m_{12}^{-1}}{m_{11}^{-1}}
% = - { m_{13} m_{11}^{-1} +  m_{23} m_{12}^{-1} \over m_{33} m_{11}^{-1}}
 = - { m_{13} (m_{22}m_{33} - m_{23} m_{32}) +  m_{23} (m_{13} m_{32}-m_{12} m_{33}) \over m_{33} m_{11}^{-1} \det m } 
  = - { m_{13} m_{22}  -  m_{23} m_{12}  \over m_{11}^{-1} \det m } \, . 
\label{22}
\end{align}
Hence, $m_{13}^{-1}/m_{11}^{-1}$ behaves as a second-order perturbation. 
The orthogonality between the first and third rows of $U_{L}$ is manifest from this expression.

%%%%%%%%%%%%%%%%%%%%%%%
\subsection{Next-to-Leading-Order corrections in SVD}
%%%%%%%%%%%%%%%%%%%%%%%

Here, we derive the next-to-leading-order (NLO) corrections to the diagonalization $U_{L}$, 
which has not been discussed in previous studies of $N_{2}$-leptogenesis. 
We will show that the leading-order (LO) results have a sufficient precision in many cases, 
because  the NLO corrections are higher-order terms of left-handed mixings or right-handed mixings suppressed by ratios of singular values. 
To describe these corrections, we introduce the anti-Hermitian generators $T_{1,2} \equiv 1 - U_{L1,2}$,  
\begin{align}
T_{1} =
\begin{pmatrix}
 0 & 0 & -\frac{m_{13}}{m_{33}} \\
 0 & 0 & -\frac{m_{23}}{m_{33}} \\
 \frac{m_{13}^*}{m_{33}^*} & \frac{m_{23}^*}{m_{33}^*} & 0 \\
\end{pmatrix} , 
~~~
T_{2} =
\begin{pmatrix}
0 & \frac{m_{12}^{-1}}{m_{11}^{-1}} & 0 \\
- \big( \frac{m_{12}^{-1}}{m_{11}^{-1} } \big)^{*} & 0 & 0 \\
 0 & 0 & 0 \\
\end{pmatrix} .
\end{align}
Some of the NLO contributions are estimated as
\begin{align}
T_{1}^{2} &=  
\begin{pmatrix}
 -\frac{| m_{13}| ^2}{| m_{33}| ^2} & -\frac{m_{13} m_{23}^*}{| m_{33}| ^2} & 0 \\
 -\frac{m_{23} m_{13}^*}{| m_{33}| ^2} & -\frac{| m_{23}| ^2}{| m_{33}| ^2} & 0 \\
 0 & 0 & -\frac{| m_{13}| ^2+| m_{23}| ^2}{| m_{33}| ^2} \\
\end{pmatrix} ,  
\label{T12} ~~~ 
T_{1}^{3} = 
\begin{pmatrix}
 0 & 0 & \frac{m_{13}}{m_{33}} \frac {| m_{13}|^2 + |m_{23}|^2}{ |m_{33}|^2} \\
 0 & 0 & \frac{m_{23}}{m_{33}} \frac{| m_{13}|^2 + |m_{23}|^2}{ |m_{33}|^2} \\
*' & *' &  0 \\
\end{pmatrix} , 
\end{align}
and 
\begin{align}
 T_{2}^{2} &= 
\begin{pmatrix}
 -\frac{| m^{-1}_{12}| ^2}{| m^{-1}_{11}| ^2} & 0 & 0 \\
 0 & -\frac{| m^{-1}_{12}| ^2}{| m^{-1}_{11}| ^2} & 0 \\
 0 & 0 & 0 \\
\end{pmatrix} , ~~~
T_{2}^{3} 
= 
\begin{pmatrix}
 0 & -\frac{m_{12}^{-1}}{m^{-1}_{11} } \frac{|m_{12}^{-1}|^2 }{|m_{11}^{-1}|^2  } & 0 \\
 \frac{(m^{-1}_{12})^*}{(m^{-1}_{11})^*} \frac{|m^{-1}_{12}|^{2}  }{  |m^{-1}_{11}|^{2} } & 0 & 0 \\
 0 & 0 & 0 \\
\end{pmatrix}  ,
\end{align}
where the matrix element denoted by $*'$ is determined by anti-Hermiticity. 
These corrections of the diagonal and off-diagonal elements start at second and third order and are suppressed by factors of $|m_{33}|^{-2}$ or $|m^{-1}_{11}|^{-2}$, the second power of left-handed mixings relative to the LO contributions. 
The only exception is the (1,2) element of Eq.~(\ref{T12}), which is also interpreted as third order once  $m_{13}/ m_{33}$  is treated as a second-order quantity from Eq.~(\ref{1}).

The remaining NLO contributions can be estimated by dividing the residual off-diagonal terms from the LO diagonalization by the corresponding singular values. 
Starting from the off-diagonal element of Eq.~(\ref{m2L1L2}),
 we neglect the smaller terms containing  $m_{12}^{-1}$. 
By extracting the largest term proportional to $m_{33}$ from $(m^\dagger m)_{23}$ and dividing it by the largest singular value $|m_{33}|^{2}$, the corresponding NLO corrections are suppressed by $|m_{33}|^{3}$ 
\begin{align}
d T^{ (2)}  \simeq
\begin{pmatrix}
 0 & 0 & 0 \\[3pt]
 0 & 0  & { m^{-1}_{11} \det m  \over m_{33}^{2} }{m^{ *}_{32} \over m_{33}^{*} }    \\[3pt]
 0 & *' & 0 \\
\end{pmatrix} . 
\label{dU2}
\end{align}
The correction to the 2-3 mixing is given by the mixing angle of the right-handed unitary matrix  $U_{R1}$~(\ref{UR}) suppressed by the ratio of the singular values in Eq.~(\ref{diag}).

Similarly, NLO corrections from $H^{1}$~(\ref{m1L1L2}) are approximately as
\begin{align}
 d T^{(1)} 
 \simeq 
\begin{pmatrix}
0 &  {m_{33} \over (m_{11}^{-1})^{2} \det  m }
  \frac{  ( m^{-1}_{21})^* }  {(m^{-1}_{11})^* } & 
{1 \over m_{33}  m^{-1}_{11}}  \frac{ m^{*}_{31} }{ m_{33}^{*}}   \\[4pt]
*' & 0 & -{ m_{21}^{-1} \det m^{*} \over  m_{33}^{2} }  { m^{*}_{31} \over m_{33}^{*} } \\[4pt]
*' &  *' &  0 \\
\end{pmatrix} . 
\end{align}
Although an additional correction arises in the 2-3 mixing,
it is of higher order than Eq.~(\ref{dU2}) from the hierarchy $|m^{-1}_{11}| \gg |m^{-1}_{21}|$ and $|m_{32}| \gg |m_{31}|$, and is not  an NLO contribution.
The corrections to the 1-2 mixing and 1-3 mixing are also the right-handed 1-2 and 1-3 mixing in Eq.~(\ref{UR}) suppressed by the ratio of the singular values,
$| m_{33}/ (m_{11}^{-1})^{2} \det  m|$ and ${1/ m_{33}  m^{-1}_{11}} $. 
These NLO structures are a consequence of the odd-order structure of the perturbative expansion, 
and are also consistent with the rephasing transformations of the left- and right-handed fields.

Finally, the NLO corrections to the singular values $\delta m_i$ 
also contribute to the NLO corrections through the LO mixing.
These $\delta m_i$ are evaluated after completing the full NLO diagonalization induced by $T_{1,2}^{2}$ and $dT^{(1,2)}$. These contributions are also suppressed at second order in the left- and right-handed mixings relative to the LO singular values. 
This fact can also be understood as a consequence of the structure of rephasing invariance.
Therefore, the qualitative conclusions remain unchanged. 

In particular, the NLO corrections to off-diagonal elements $U_{L ij}$ are third-order perturbations of mixing angles of $U_{L}$, or right-handed mixing angles in $U_{R1,2}$ suppressed by ratios of singular values. 
Although the right-handed mixings are, in principle unknown, one can always redefine the right-handed fields to choose a basis in which the mixings are sufficiently small. Therefore, by working in an appropriate basis, 
the NLO contributions are suppressed by the square of the Cabibbo angle $\lambda^2 \sim O(4 \%)$ relative to the LO terms. 
In most situations, the LO diagonalization already achieves an accuracy comparable to the current experimental precision $O(1 \%)$. 

%%%%%%%%%%%%%%%%%
\section{Perturbative Evaluation of CP Phase using Mass Matrices}
%%%%%%%%%%%%%%%%%

Using the perturbative SVD, we represent the mixing angles and CP phase $\delta$ in the CKM matrix.  
For the quark mass matrices $m_{u,d}$ in the SM Lagrangian,
\begin{align}
\Lg \ni  \sum_{f} -  \bar f_{Li } m_{f ij} f_{Rj} + {\rm h.c.} \, , 
\end{align}
diagonalization of the mass matrices $ m_{f} = U_{Lf}^{\dg} m_{f}^{\rm diag} U_{Rf} $ 
leads to the following CKM  matrix 
\begin{align}
V_{\rm CKM}  = U_{L u }^{} U_{Ld}^{\dg}  \, . 
\end{align}
With the perturbative diagonalization defined analogously for the up- and down-quark masses, 
the perturbative CKM matrix is given by
\begin{align}
V_{\rm CKM} \simeq 
\begin{pmatrix}
1 &
-\frac{m^{-1}_{d12}}{m^{-1}_{d11}} + \frac{m^{-1}_{u12}}{m^{-1}_{u11}} & 
\frac{m_{d13}}{m_{d33}}+\frac{m_{d23}^{}}{m_{d33}^{}} \frac{m^{-1}_{u12}}{ m^{-1}_{u11}} + \frac{m^{-1}_{u13}}{m^{-1}_{u11}} \\[4pt]
 \big ( \frac{m^{-1}_{d12}}{m^{-1}_{d11}} - \frac{m^{-1}_{u12}}{m^{-1}_{u11}} \big )^{*} &
 1 &
\frac{m_{d23}}{m_{d33}} - \frac{m_{u23}}{m_{u33}} \\[6pt]
\big ( \frac{m_{u13}}{m_{u33}} +\frac{ m_{u23}}{m_{u33} } \frac{ m^{-1}_{d12} }{ m^{-1}_{d11} } + \frac{ m^{-1}_{d13} }{ m^{-1}_{d11} }  \big )^{*} &
-\frac{m_{d23}^*}{m_{d33}^*}+\frac{m_{u23}^*}{m_{u33}^*} & 
 1 \\
\end{pmatrix} ,
\label{mCKM}
\end{align}
where NLO corrections are suppressed at the second order. 

Thus, the three mixing angles are represented by rephasing invariants 
\begin{align}
 \tan^{2} \th_{12} &= {V_{u s}^{*} V_{u s}^{} \over V_{u d}^{*} V_{ud}^{} } 
 \simeq \left | \frac{m^{-1}_{u12}}{m^{-1}_{u11}} -\frac{m^{-1}_{d12}}{m^{-1}_{d11}} \right |^{2} ,  \label{mixing1}\\
 \tan^{2} \th_{23} &= {V_{c b}^{*} V_{c b}^{} \over V_{t b}^{*} V_{t b}^{} } 
 \simeq \left | \frac{m_{d23}}{m_{d33}} - \frac{m_{u23}}{m_{u33}} \right |^{2}  , \\ 
 \sin^{2}  \th_{13} &= V_{ub}^{*} V_{ub} \simeq \left | \frac{m_{d13}}{m_{d33}}+\frac{m_{d23}^{}}{m_{d33}^{}} \frac{m^{-1}_{u12}}{ m^{-1}_{u11}} + \frac{m^{-1}_{u13}}{m^{-1}_{u11}} \right|^{2} . 
 \label{mixing3}
\end{align}
The CP phase of the mixing matrix in the PDG parametrization is evaluated 
by a rephasing invariant formula \cite{Yang:2025hex}
\begin{align}
\d = \arg \left[ {V_{ud} V_{us} V_{cb} V_{tb}  \over V_{ub} \det V_{\rm CKM}} \right] \, . 
\end{align}
Although these matrix elements and the determinant transform under rephasing, this entire set is rephasing invariant and corresponds to a physical observable. 
Since the rephasing invariants take the same values in any phase convention,
we can evaluate  the quantities $\theta_{ij}$ and $\delta$ in a specific convention. 
This phase $\delta$ coincides with the value  derived from the Jarlskog invariant $J$ \cite{Yang:2025law}
\begin{align}
\sin \arg \left [  {V_{ud} V_{us} V_{cb} V_{tb}  \over V_{ub} \det V_{\rm CKM}} \right ]  =  \frac{  1 - |V_{ub}^{2}| }{ |V_{ud} V_{us} V_{c b} V_{tb} V_{ub}| } J 
= \sin \d \, . 
\end{align}
Compared with the invariant $J$, the expression has several advantages: 
(i) it can be decomposed using the identity $\arg[ab]=\arg a+\arg b$ which is simpler than the addition theorem of $\sin (a + b)$, 
(ii) this phase contains the signature of $\cos \d$, and finally  
(iii) its constraint is independent of those of the mixing angles~(\ref{mixing1})-(\ref{mixing3}). 

From the perturbative expansion, we obtain the LO expression for the CP phase. Since we work in the basis with $\det V_{\rm CKM}=1 + O(m_{ii}^{2} / m_{ij}^{2})$, the CP phase $\d$ is found to be
\begin{align}
\d = \arg \left[ 
{\left (  \dfrac{m^{-1}_{u12}}{m^{-1}_{u11}} - \dfrac{m^{-1}_{d12}}{m^{-1}_{d11}} \right )
\left ( \dfrac{m_{d23}}{m_{d33}}-\dfrac{m_{u23}}{m_{u33}}  \right )
 \over 
\dfrac{m_{d13}}{m_{d33}}+\dfrac{m_{d23}^{}}{m_{d33}^{} } \dfrac{m^{-1}_{u12}}{m^{-1}_{u11}}+\dfrac{m^{-1}_{u13}}{m^{-1}_{u11}} }  \right ]  + O \lsp {m_{fij}^{2} \over m_{fii}^{2}} \rsp \, . 
\end{align}
In the basis where $m_{u}$ is diagonal, 
the CP phase reduces to 
\begin{align}
\d \simeq \arg \left[  - \frac{ m^{-1}_{d12}}{ m^{-1}_{d11}} \frac{m_{d23}^{} }{m_{d13}^{} } \right ]  \, ,
\end{align}
a fourth-order invariant constructed from the mass matrix and its inverse. 

%%%%%%%%%%%%%%%%%%%%%%%%%%%%%%%%%%%
\subsection{CP phase in Kobayashi--Maskawa parametrization}
%%%%%%%%%%%%%%%%%%%%%%%%%%%%%%%%%%%

In the original Kobayashi--Maskawa parametrization, the CP phase $\d_{\rm KM}$ is known to be nearly maximal $\d_{\rm KM} \simeq \pi/2$ 
\cite{Koide:2004gj, Koide:2008yu, Hocker:2006xb, Frampton:2010ii, Dueck:2010fa, Frampton:2010uq,Li:2010ae,Qin:2011bq,Zhou:2011xm,Qin:2011ub,Qin:2010hn, Zhang:2012ys, Li:2012zxa, Zhang:2012bk}. 
The parameters of the CKM matrix from the latest UTfit are given by \cite{UTfit:2022hsi} 
\begin{align}
\sin \th_{12} &= 0.22519 \pm 0.00083 \, ,  ~~~ \sin \th_{23} = 0.04200 \pm 0.00047 \, , \nn \\
\sin \th_{13} &= 0.003714 \pm 0.000092 \, ,  ~~~ \d = 1.137 \pm 0.022  = 65.15^{\circ} \pm1.3^{\circ}  \, . 
\end{align}
From the best fit values, the CP phase $\d_{\rm KM}$ can likewise be evaluated as \cite{Yang:2025ftl}
\begin{align}
\d_{\rm KM } = \arg \left[ - { V_{ud} \det V_{\rm CKM} \over  V_{us} V_{ub} V_{cd} V_{td} } \right] = 87.56^{\circ} \, . 
\label{obsdkm}
\end{align}

Due to the unitarity, one finds $\arg [V_{us} V_{cd}] \simeq -1$ in Eq.~(\ref{mCKM}) and 
the CP phase is reduced to
\begin{align}
\d_{\rm KM}
 %& =  \arg \left[ - { V_{ud} \det V_{\rm CKM} \over  V_{us} V_{ub} V_{cd} V_{td}} \right] 
= \arg \left[ + { V_{td}^{*} \over  V_{ub} } \right] + O \lsp {m_{fij}^{2} \over m_{fii}^{2}} \rsp
 \simeq \arg \left[  
   {\dfrac{m_{u13}}{m_{u33}} + \dfrac{ m_{u23}^{} }{m_{u33}^{} } \dfrac{m^{-1}_{d12}}{m^{-1}_{d11}} + \dfrac{m^{-1}_{d13}}{m^{-1}_{d11}}  \over 
\dfrac{m_{d13}}{m_{d33}} + \dfrac{m_{d23}^{}}{m_{d33}^{} } \dfrac{m^{-1}_{u12}}{m^{-1}_{u11}} + \dfrac{m^{-1}_{u13}}{m^{-1}_{u11}}  }  \right ]  \, . 
\end{align}
As an illustrative example, let us consider the limit in which the 1-3 mixings, 
$m_{(u,d)13}/ m_{(u,d)33}$ are sufficiently small or exact zero textures. 
Since the ratio $m_{13}^{-1} / m_{11}^{-1} $ is reduced to the product of the two mixing angles $-\frac{m_{12}^{-1}}{m_{11}^{-1}} \frac{m_{23}}{m_{33}}$ in Eq.~(\ref{21}),
\begin{align}
\d_{\rm KM} & \simeq  
\arg \left[ { \dfrac{m_{d12}^{-1}}{m_{d11}^{-1}} 
\bigg ( \dfrac{ m_{u23} }{m_{u33} } -  \dfrac{m_{d23}}{m_{d33}}  \bigg ) \over 
\dfrac{m_{u12}^{-1}}{m_{u11}^{-1}}
\bigg ( \dfrac{m_{d23}}{m_{d33}} - \dfrac{m_{u23}}{m_{u33}} \bigg )  }  \right ]  
 = 
\arg \left[ - { m^{-1}_{u11}  \over  m^{-1}_{u12}  }  { m^{-1}_{d12} \over m^{-1}_{d11}  } \right ] 
\simeq {\pi \over 2} \, . 
\end{align}
It corresponds to the well-known texture that the relative phase between the first and second generations is nearly maximal \cite{Shin:1985cg, Gronau:1985tx, Fritzsch:1985yv, Kang:1985nw,  Lehmann:1995br, Kang:1997uv, Fritzsch:1999im, Xing:2003yj, Antusch:2009hq,Tanimoto:2015hqa, Yang:2020qsa, Yang:2020goc, Yang:2021smh}, 
\begin{align}
m_{u} = \Diag{i}{1}{1}
\begin{pmatrix}
m_{u 11}' & m_{u 12}' & 0 \\
m_{u 21}' & m_{u 22}' & m_{u 23}' \\
m_{u 31}' & m_{u 32}' & m_{u 33}' \\
\end{pmatrix} , 
~~~
m_{d} = 
\begin{pmatrix}
 m_{d11}' &  m_{d12}' & 0 \\
 m_{d21}' &  m_{d22}' & m_{d23}' \\
 m_{d31}' &  m_{d32}' & m_{d33}' \\
\end{pmatrix} , 
\end{align}
where all the parameters $m_{u ij}'$ and $m_{d ij}'$ are real. 

Alternatively, in the basis where $m_{u}$  is diagonal, 
\begin{align}
 \d_{\rm KM} \simeq \arg \left [ 
\frac{ m^{-1}_{d13} }{ m^{-1}_{d11}} 
 \frac{m_{d33}^{} }{m_{d13}^{} } \right ] \, .  
\end{align}
Such a texture satisfying this condition can be realized by
attributing the up-type quark mixing to the down-type sector 
\begin{align}
m_{d} & =
\begin{pmatrix}
1 & (m_{u}')^{-1}_{12} / (m_{u}')^{-1}_{11} & 0 \\
- (m_{u}')^{-1}_{12} / (m_{u}')^{-1}_{11} & 1 & 0 \\
0 & 0 & 1
\end{pmatrix}
\Diag{i}{1}{1}
\begin{pmatrix}
 m_{d11}' &  m_{d12}' & 0 \\
 m_{d21}' &  m_{d22}' & m_{d23}' \\
 m_{d31}' &  m_{d32}' & m_{d33}' \\
\end{pmatrix} \, ,  \\
m_{d}^{-1} & =
\begin{pmatrix}
 m_{d11}' &  m_{d12}' & 0 \\
 m_{d21}' &  m_{d22}' & m_{d23}' \\
 m_{d31}' &  m_{d32}' & m_{d33}' \\
\end{pmatrix}^{-1}
\Diag{-i}{1}{1}
\begin{pmatrix}
1 & - (m_{u}')^{-1}_{12} / (m_{u}')^{-1}_{11} & 0 \\
(m_{u}')^{-1}_{12} / (m_{u}')^{-1}_{11} & 1 & 0 \\
0 & 0 & 1
\end{pmatrix}
 \, . 
\end{align}
In this case, $m_{d11}^{-1}$ becomes purely imaginary at LO, implying that
$\delta_{\rm KM} \simeq \pm \pi/2$. 

The appearance of the inverse mass matrix in the description of the CP phase motivates the use of the adjugate approach~\cite{Denton:2019pka,Abdullahi:2022fkh}. Although this method allows the CP phase to be expressed as an exact function of the mass matrix, the resulting expressions become excessively complicated, making it difficult to obtain a practically useful formula.

%%%%%%%%%%%%%%%%%%%%
\section{Conclusions}
%%%%%%%%%%%%%%%%%%%%

In this paper, we derive approximate expressions for
the CP phase $\d$ in the CKM matrix for hierarchical mass matrices of quarks.  
By successively integrating out the heavier generations through a seesaw-like procedure, 
the diagonalization is perturbatively expressed in terms of the mass matrices $m_{u,d}$ and their inverses $m_{u,d}^{-1}$.
Using a rephasing invariant formula, the CP phase in the basis where $m_{u}$ is diagonal is reduced to a fourth-order rephasing invariant 
$\d \simeq \arg [  - {m^{-1}_{d12} m_{d23}^{} /  m^{-1}_{d11} m_{d13}^{} } ] $. 
%constructed from the down-quark mass matrix and its inverse. 
%
Furthermore, in the same basis, the CP phase of the Kobayashi--Maskawa parametrization is likewise expressed as an invariant  $\d_{\rm KM} \simeq \arg  [ { m^{-1}_{d13} m_{d33}^{} / m^{-1}_{d11} m_{d13}^{} }] \simeq \pi /2$.

We also systematically evaluate the NLO corrections and show that they are either higher-order terms of left-handed mixings or right-handed mixings suppressed by ratios of singular values. 
Consequently, in a basis where the right-handed mixings are sufficiently small, all the NLO corrections are suppressed by the second order $\lambda^{2} \sim O(4\%)$ relative to the leading-order terms. 
Therefore, the leading-order approximation  already achieves an accuracy comparable to the current experimental precision of $O(1\%)$.

These results establish a direct correspondence between the observed CP phases and the underlying structure of hierarchical quark mass matrices, and represent one of the essential aspects of CP violation in the quark sector. 
Besides, rephasing invariants constructed from inverse matrices remain largely unexplored, and a systematic investigation of this new class of invariants will be useful for flavor models with texture zeros and grand unified theories. 
It will also provide new insights into residual symmetries~\cite{Ge:2011qn,Petcov:2014laa,Novichkov:2018yse,Ge:2025csr,Dutta:2026dzh}, generalized CP symmetries~\cite{Ecker:1983hz,Gronau:1985sp,Feruglio:2012cw,Holthausen:2012dk}, and the strong CP problem, for investigating the origin of CP violation.

%\bibliographystyle{bib/h-physrev50}
%\bibliography{bib/fourzero, bib/onezero, bib/refsym, bib/mutausym,bib/PSGUT,bib/StrongCP,bib/LR,bib/GCP,bib/U(2),bib/flaxion,bib/minimal-natural,bib/chiral, bib/T2HK,bib/CKM2MNS,bib/KMCPV,bib/TM12,bib/N2leptogenesis}

\end{document}